\documentclass[prb,twocolumn,floatfix]{revtex4}
\usepackage{graphicx}
\begin{document}
\title{Energy diffusion in frustrated quantum spin chains exhibiting 
Gaussian orthogonal ensemble level statistics}

\author{Kazue Kudo}
\affiliation{Graduate School of Humanities and Sciences, Ochanomizu
University, 2-1-1 Ohtsuka, Bunkyo-ku, Tokyo 112-8610, Japan}
\email{kudo@degway.phys.ocha.ac.jp}
\altaffiliation[Present address: ]{Department of Applied Physics, 
Osaka City University, Osaka 558-8585, Japan}
\author{Katsuhiro Nakamura}
\affiliation{Department of Applied Physics, Osaka City University, Osaka
558-8585, Japan}
\email{nakamura@a-phys.eng.osaka-cu.ac.jp}

\date{\today}

\begin{abstract}
Frustrated quantum $XXZ$ spin chains with the
     next-nearest-neighbor  (NNN) couplings are typically deterministic 
many-body
     systems exhibiting Gaussian orthogonal ensemble (GOE) spectral statistics.
We investigate
energy diffusion for these spin chains in the presence of
a periodically oscillating magnetic
field. Diffusion coefficients are found to obey the power law with respect to
both the field strength and driving frequency with its power varying
depending on the
linear response and non-perturbative regimes. 
The widths of the linear response and the non-perturbative regimes
 depend on the strength of frustrations. 
We have also elucidated a mechanism for oscillation of energy diffusion
in the case of weakened frustrations.
\end{abstract}


\maketitle

\section{\label{sec:intro} Introduction}

There exists an accumulation of studies on quantum dynamics of classically
chaotic systems
, e.g.,   kicked rotators,  kicked spin-tops, hydrogen atoms in
time-dependent electric field,
and the standard map
model, to mention a few.~\cite{NS2001}
 Quantum suppression of energy diffusion,
dynamical localization
and other signatures of quantum chaos
are notable in these dynamics. However, most of the systems treated so far are
confined to those with a few degrees-of-freedom, and little attention is
paid to dynamics of quantum many-body systems\cite{Jona,Prosen,Flambaum}
 whose adiabatic energy levels
are characterized by
Gaussian orthogonal ensemble (GOE) spectral statistics, i.e., by a hallmark
of quantum chaos.
While some important
contributions~\cite{Wilkinson88,WilAus92,WilAus95,Bulgac,Cohen00D,Cohen00M,Machida}
are devoted to dynamics of a kind of many-body systems,
those systems are actually described by the random-matrix models,
and not by deterministic quantum Hamiltonians.
It is highly desirable to explore dynamical behaviors of deterministic
quantum many-body systems
     exhibiting GOE or GUE spectral statistics.
 
     On the other hand, the frustrated quantum spin systems have been receiving
     a wide attention, and  we can find their realization in $s=\frac12$
     antiferromagnetic chains Cu(ampy)Br$_2$~\cite{Kikuchi} and
     (N$_2$H$_5$)CuCl$_3$,~\cite{Hagiwara} and in $s=\frac12$ triangular
     antiferromagnets.~\cite{KPL}
     The high-lying states of these quantum many-body systems deserve
     being   studied
     in the context of "quantum chaos." The advantage of the frustrated quantum
systems
     is that one can expect quantum chaotic behaviors
     appearing already in the low energy region
     near the ground state.~\cite{Nakamura85,Yamasaki04}
    From the viewpoint of real physics of condensed matters, novel features
observed
in the low-energy region are very important and welcome.
Recalling that in most of deterministic Hamiltonian systems quantum chaotic
behaviors
appear in high-lying states, the role of
frustration is essential in the study of
quantum dynamics from the ground state of deterministic many-body systems with
GOE or Gaussian unitary ensemble (GUE) level statistics.

In this paper, we investigate dynamics of
$XXZ$ quantum spin chains which have antiferromagnetic exchange
interactions for the
nearest-neighbor (NN) and
the next-nearest-neighbor (NNN) couplings.  The NNN couplings cause the
frustration, i.e., difficulty in achieving the ground state,
thereby attributing a name of frustrated quantum spin chains to these systems.
In fact, the level statistics of the NNN coupled $XXZ$ spin chains
without an applied magnetic field
has been studied intensively in Refs.~\onlinecite{Kudo03,Kudo04}, and
it has been shown that GOE behavior,
which is typical of quantum chaos, appears already in the low energy
region near the ground state.~\cite{note,note1}
The ground-state phase diagram is shown in Ref.~\onlinecite{Nomura} for
the NNN coupled $XXZ$ spin chains without a magnetic field.

A natural extension of the research is to investigate dynamics
of the frustrated quantum spin chains
with an applied periodically oscillating
magnetic field.  We calculate a time evolution of the system
starting from their ground state and analyze
the nature of energy diffusion. We shall numerically exhibit
the time dependence of energy variance,
and show how
the diffusion coefficients depend on the coupling constants, the anisotropy
parameters, the magnetic field and the frequency of the
field.  Furthermore, to compare with the energy diffusion in
the case of weakened frustrations, we also investigate dynamics of
the corresponding energy diffusion in $XXZ$ spin chains with
small NNN couplings.
 
The organization of the paper is as follows:
In Sec.~\ref{sec:method}, we briefly describe a numerical approach to
obtain the time evolution operator. In Sec.~\ref{sec:variance}
we shall show the time dependence of energy variance starting from
the ground state of the many-body system and explain a way to evaluate
diffusion coefficients. Section \ref{sec:diffusion} elucidates
how diffusion coefficients depend on field strength and driving
frequency. 
Here 
power laws are shown to exist in the linear response and non-perturbative
regions.
Section \ref{sec:compare} is devoted to a mechanism of oscillation of energy
diffusion. 
Conclusions are given in Sec.~\ref{sec:conc}.

\section{\label{sec:method} Numerical Procedure}

We give Hamiltonian for the NN and NNN exchange-coupled
spin chain on $L$ sites with a time-periodic oscillating magnetic field as
\begin{equation}
\mathcal{H}(t)=\mathcal{H}_0 +\mathcal{H}_1(t),
\label{eq:H}
\end{equation}
where
\begin{eqnarray}
     \mathcal{H}_0 &=& J_1\sum_{j=1}^{L}(S^x_j S^x_{j+1} +S^y_j S^y_{j+1}
     +\Delta S^z_j S^z_{j+1})
\nonumber \\
     &+& J_2\sum_{j=1}^{L}(S^x_j S^x_{j+2} +S^y_j S^y_{j+2}
     +\Delta S^z_j S^z_{j+2})   \nonumber \\
&-& \sum_{j=1}^{L}B^z_j(0)S^z_j,
\end{eqnarray}
\begin{equation}
    \mathcal{H}_1(t)= \sum_{j=1}^{L}B^z_j(0)S^z_j -\sum_{j=1}^{L}B^z_j(t)S^z_j.
\end{equation}
Here, $S_j^{\alpha}=(1/2)\sigma_j^{\alpha}$ and 
$(\sigma^x_j, \sigma^y_j, \sigma^z_j)$ are the Pauli matrices on the
$j$th  site;
the periodic boundary conditions (P.~B.~C.) are imposed.
The magnetic field $B^z_j$ on $j$th site along the $z$ axis is chosen  to
form a traveling wave:
\begin{equation}
    B^z_j(t)=B_0\sin\left( \omega t-\frac{2\pi j}L\right).
\label{eq:jiba}
\end{equation}
The period of Eq.~(\ref{eq:H}) as well as Eq.~(\ref{eq:jiba}) is
 $T=2\pi/\omega$. Because of the coexisting spatial P.~B.~C., however,
 the effective  period of the adiabatic energy spectra is
 given by $T'=T/L=2\pi/(\omega L)$. In other words, the period of the
Hamiltonian operator is $T$, and the spectral flow of the
 eigenvalues has the effective period $T'$.
This periodicity property comes from the
 traveling-wave form of Eq.~(\ref{eq:jiba}), and is advantageous for our
 getting a sufficient number of relevant data in each period $T$.

When $J_1>0$ and $J_2>0$, 
the unperturbed  Hamiltonian $\mathcal{H}_0$
without coupling to the magnetic field is translationally invariant and 
corresponds to a
frustrated antiferromagnetic quantum spin model
exhibiting GOE level statistics.~\cite{Kudo03,Kudo04} 
If $J_2=0$ and $B_0=0$, it describes an
integrable  and non-frustrated model. Before calculating energy
diffusion, we have to consider the symmetries of the model. We divide
the Hamiltonian matrix to some sectors which have the same quantum
numbers. In the Hamiltonian Eq.(\ref{eq:H}), total $S^z $ $(S^z_{\rm tot})$ is
conserved. The eigenstates with different $S^z_{\rm tot}$ are
uncorrelated.
On the other hand, the non-uniform magnetic field
breaks the translational symmetry, and leads
to mixing between manifolds of different wave-number values.

Before proceeding to consider the time evolution of a wave
function, we should note: If we use the original
Hamiltonian
$\mathcal{H}(t)=\mathcal{H}_0 +\mathcal{H}_1(t)$ as it
stands, the mean level spacing of eigenvalues would change
depending
on $J_2$, $\Delta$, and $B_0$.
To see a universal feature of the energy diffusion, it is
essential to
scale the Hamiltonian so that the full range of adiabatic
energy eigenvalues becomes almost free from these
parameters.
Noting that this energy range for the original Hamiltonian
is 
of order of $L$  when $J_1=J_2=\Delta=1$,
we define the scaled Hamiltonian  $H(t)=H_0+H_1(t)$
so that the full energy range equals $L$ at $t=0$, 
which will be used throughout in the text.
The Sch\"odinger equation is then given by
\begin{equation}
    i\hbar \frac{\partial}{\partial t}|\psi (t)\rangle
    =H(t)|\psi (t)\rangle
    =[H_0+H_1(t)] |\psi (t)\rangle.
\label{eq:Schrodinger}
\end{equation}
The solution of Eq. (\ref{eq:Schrodinger}) consists
of a sequence of the infinitesimal processes as
\begin{eqnarray}
     |\psi (t)\rangle &=& U(t;t-\Delta t) U(t-\Delta t;t-2\Delta t)
\nonumber \\
&\cdots& U(2\Delta t;\Delta t)
    U(\Delta t;0) |\psi (0)\rangle.
\end{eqnarray}
The initial state $|\psi(0)\rangle$ is taken to be the ground state,
since our concern lies in the dynamical behaviors starting from the
many-body ground state.
To calculate a time evolution operator $U(t+\Delta t;t)$
for each short time step $\Delta t$, we use the
fourth-order decomposition formula for the exponential
operator:~\cite{Suzuki90}
\begin{eqnarray}
    U(t+\Delta t;t)&=& S(-i p_5\Delta t/\hbar,t_5)
S(-i p_4\Delta t/\hbar,t_4) \nonumber\\
&\cdots& S(-i p_2\Delta t/\hbar,t_2)
                       S(-i p_1\Delta t/\hbar,t_1),
\label{eq:U}
\end{eqnarray}
where,
\begin{equation}
    S(x,t)=\exp\left( \frac{x H_1(t)}2 \right) \exp(x H_0)
\exp\left( \frac{x H_1(t)}2 \right).
\end{equation}
Here, $t_j$'s and $p_j$'s are the following:
\begin{eqnarray}
    t_j &=& t+(p_1+p_2+\cdots +p_{j-1}+p_j/2)\Delta t, \nonumber\\
    p &=& p_1 =p_2 =p_4=p_5, \nonumber\\
      &=& 0.4144907717943757\cdots \nonumber\\
    p_3&=&1-4p.
\end{eqnarray}
The numerical procedure based on the above decompositions is quite effective
when $H_1(t)$ and $H_0$
do not commute and each time step is very small. 
Our computation below is concerned mainly with
the system of $L=10$, whose $S^z_{\rm tot}=1$ manifold involves 210
levels. To check the validity of our assertion, some of the results will be
compared to those for the system of $L=14$ and 
$S^z_{\rm tot}=4$ whose manifold involves 364 levels.

\section{\label{sec:variance} Time Dependence of Energy Variance}

\begin{figure}
\includegraphics[width=8cm]{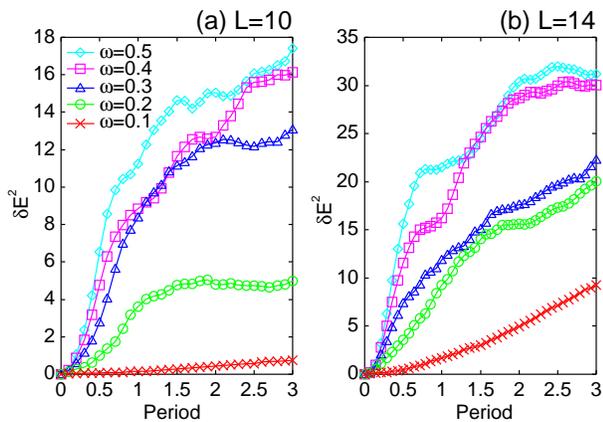}
\caption{\label{fig:1} (Color online) 
Time evolution of energy diffusion for (a) $L=10$ and (b) $L=14$. The
    parameters are the following: $J_1=J_2=1.0$,
    $\Delta=0.3$,  $B_0=1.0$.}
\end{figure}

We calculate time evolution of the state and evaluate
energy variances at each integer multiple of
the effective period $T'=T/L=2\pi /(\omega L)$.
As mentioned already, we choose the ground state as an initial state,
following the spirit of real physics of condensed matters.
This viewpoint is in contrast to that of
the random matrix models where initial states
are chosen among high-lying
ones.~\cite{Wilkinson88,WilAus92,WilAus95,Bulgac,Cohen00D,Cohen00M} 
Consequently,
the energy variance of our primary concern is
the \textit{variance around the ground state energy} $E_0$
and is defined by
\begin{equation}
    \delta E(t)^2= \langle \psi (t)|[H(t)-E_0]^2 |\psi (t) \rangle .
\label{eq:variance}
\end{equation}
Time evolution of $\delta E(t)^2$ is shown in Fig.~\ref{fig:1}. The
parameters except for $\omega$ are fixed. The larger $\omega$ is, the faster
the energy diffusion grows, which is consistent with our expectations. The
details will be explained in Sec.~\ref{sec:diffusion}.
For wide parameter values of the next-nearest-neighbor (NNN) coupling $J_2$
and exchange anisotropy $\Delta$, the early stage of quantum dynamics
becomes to show the normal diffusion in energy space, i.e., a linear growth of 
$\delta E(t)^2$ in time.
While we proceed to investigate this normal diffusion process, 
energy variances will
finally saturate because the system size we consider is finite. On the
other hand, energy variances can also saturate because of another reason,
i.e., the dynamical localization effect associated with a periodic
perturbation. 

During the first period, $\delta E(t)^2$ shows a linear
growth in time as shown in Fig.~\ref{fig:1} (a). The range of the linear
growth is not sufficiently wide because the number of levels is not
large enough for
$L=10$. However, if the number of levels as well as the system
size is increased, the length of a linear region may be elongated. 
In fact, the linear growth of $\delta E(t)^2$ during the first
period can be recognized more clearly for $L=14$ than for $L=10$ 
[see Fig.~\ref{fig:1} (b)].
The diffusion coefficient has to be determined much earlier than the 
time where saturation begins. 
We determine the diffusion coefficient $D$  from the fitting
\begin{equation}
    \delta E(t)^2 = Dt +\mbox{\rm const.}
\label{eq:defD}
\end{equation}
to some data points around the largest slope
in the first period, where the normal diffusion is expected. 

\section{\label{sec:diffusion} Diffusion coefficients: dependence on field
strength and frequency}

Since the time evolution of our system starts from the ground state,
we consider non-adiabatic regions where inter-level transitions
frequently occur. In other words, we suppress a near-adiabatic or the
so-called Landau-Zener (LZ) region
where the driving frequency $\omega$ is much smaller than the mean level
spacing divided by Planck constant.  
Because of a large energy gap between the ground and first
excited states,
the near-adiabatic region cannot result in the notable energy diffusion and
will be left outside a scope of the present study.

Beyond the LZ region, however, so long as the changing rate $\dot{X}$  of a
perturbation
parameter is not very large,~\cite{note2} the diffusion coefficient can be calculated
using the Kubo formula. We call such a parameter regime ``linear
response'' regime. In the linear response regime, $D\propto\dot{X}^2$
(See, e.g., Refs.~\onlinecite{WilAus92} and \onlinecite{WilAus95}).
When $\dot{X}$ is large, however, the perturbation theory fails. We call
such a parameter regime ``non-perturbative'' regime. In the 
non-perturbative regime, the diffusion coefficient is smaller than that
predicted by the Kubo formula.~\cite{WilAus95,Cohen00D} According to
Ref.~\onlinecite{WilAus95}, $D \propto \dot{X}^{\gamma}$ with
$\gamma \le 1$ in the non-perturbative regime. We note that
$\dot{X}\propto B_0\omega$ in this paper since the perturbation is given by
Eq.~(\ref{eq:jiba}). 
Both Refs.~\onlinecite{WilAus95} and \onlinecite{Cohen00D} are based on
the random matrix models, which are utterly different from our
deterministic one.

\begin{figure}
\includegraphics[width=8cm]{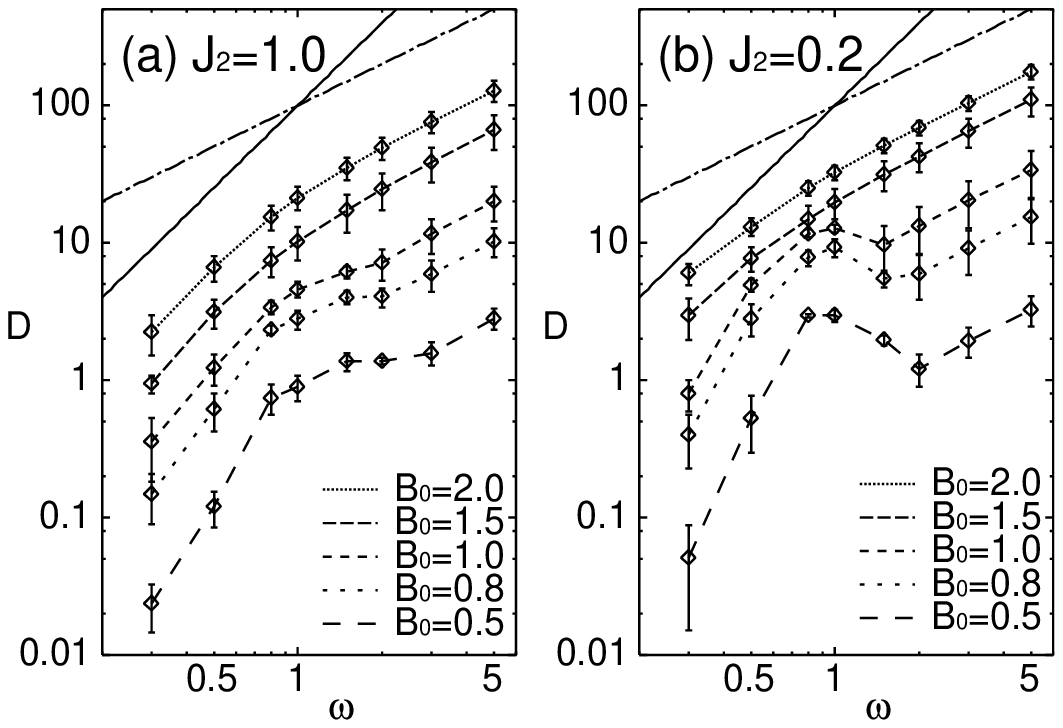}
\caption{\label{fig:2} Driving frequency dependence of the diffusion
    coefficients.  The chained line and the solid line
 are just eye guides for $D\propto \omega^{\beta}$ with $\beta=1$ and
 $2$, respectively.  
The symbols ($\diamond$) are the average of the  diffusion coefficients
    calculated for several values of $\Delta$ ($0.3\le \Delta \le 0.8$). The
    parameters are the following: $L=10$, $J_1=1.0$; (a) $J_2=1.0$, 
(b) $J_2=0.2$.}
\end{figure}

\begin{figure*}
\includegraphics[width=12cm]{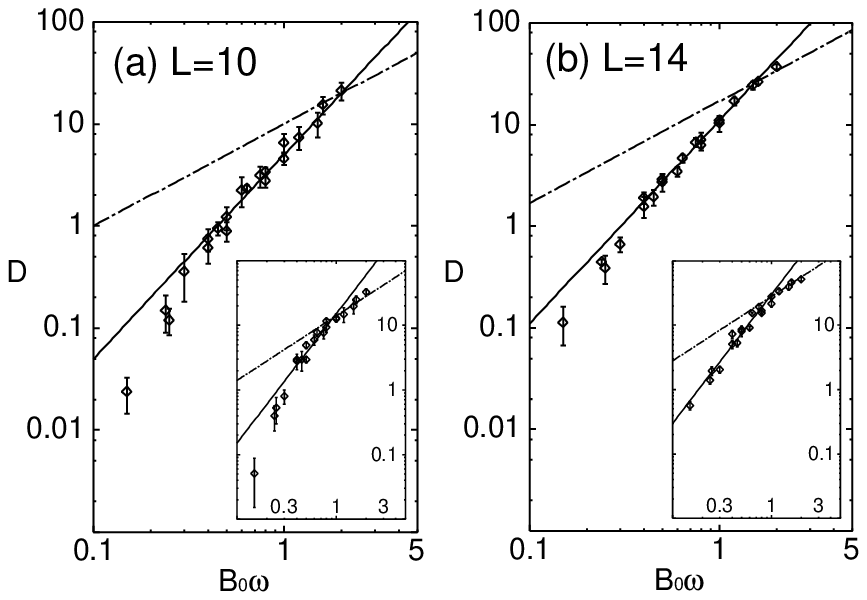}
\caption{\label{fig:3} Dependence of the diffusion
    coefficients on the product of field strength $B_0$ and driving
 frequency $\omega$ for (a) $L=10$ and (b) $L=14$. 
The symbols ($\diamond$) are the average of the  diffusion coefficient
    calculated for several values of $\Delta$ ($0.3\le \Delta \le 0.8$). 
The parameters are $J_1=J_2=1.0$; for
 the inset, $J_1=1.0$ and $J_2=0.2$. 
The chained line and the solid line
 are just eye guides for $D\propto (B_0\omega)^{\beta}$ with $\beta=1$
 and $2$, respectively. Some error bars are too short to see.}
\end{figure*}

Numerical results of diffusion coefficients in Fig.~\ref{fig:2}
 are almost consistent with the argument of Ref.~\onlinecite{WilAus95}.
Diffusion coefficients as a function of $\omega$
are shown in Fig.~\ref{fig:2}.
In Fig.~\ref{fig:2}(a), where $J_2=1.0$
(i.e., the fully-frustrated case), $D$ is larger
as $B_0$ is larger for a fixed value of $\omega$. In a small-$\omega$ regime,
$D\propto \omega^{\beta}$ with $\beta=2$, though $\beta >2$ for small $B_0$. 
The latter is merely attributed to the fact that the
perturbation is
too  small to observe a sufficient energy diffusion 
when both $\omega$ and $B_0$ are small.  
In a large-$\omega$ regime, $\beta=1$.
Namely, we observe that $\beta=2$ in the linear response
regime
and $\beta=1$ in the non-perturbative regime.
In fact, for a large-$\omega$ regime, 
the increase of energy variances per effective
period hardly depend on $\omega$ by the time when $\delta E(t)^2$
 starts to decrease.
This explains the observation
that $D\propto \omega^{\beta}$ with $\beta=1$ in both
Fig.~\ref{fig:2}(a) and Fig.~\ref{fig:2}(b). 
Let us represent the increase of energy variances per
effective period
as $\Delta(\delta E^2)$. From the definition of $D$,
i.e. Eq.~(\ref{eq:defD}), $D\propto \Delta(\delta E^2)/T'$.
If
$\Delta(\delta E^2)$ is constant, $D\propto \omega$.

On the other hand, in Fig.~\ref{fig:2}(b) where $J_2=0.2$
(i.e., a weakly-frustrated
 case),  the region with $\beta=1$ is expanding. 
For small $B_0$, $\beta>2$ in a small-$\omega$ regime is
the same as in the
case of $J_2=1.0$.
For small $B_0$ and around $\omega\sim 1$,
$D$ seems to rather decrease than increase 
especially in the case of $J_2=0.2$.
Some kind of localization would have occurred 
in the very early stage of energy diffusion for large
$\omega$ and small $B_0$,
leading to the suppression of $D$.

It is seen more clearly in
Fig.~\ref{fig:3} how the behavior of $D$ changes between a linear
response regime and a non-perturbative regime. 
The diffusion coefficient $D$ obeys the power law
$D\propto (B_0 \omega)^{\beta}$ with its power $\beta$ being two in  the
linear response regime and $\beta=1$ in the non-perturbative regime.
For small $B_0\omega$, the power law seems to fail because of some
finite-size effects. 
These universal feature is confirmed in systems of larger size.
Actually, $D$ obeys the power law 
better for $L=14$ [Fig.~\ref{fig:3}(b)] than $L=10$
[Fig.~\ref{fig:3}(a)]. In addition, error bars are shorter for $L=14$
than $L=10$.
Here, we have used the data of
$\omega \le 1$. We cannot expect meaningful results in a large-$\omega$
regime since, as mentioned above, energy diffusion is not normal there.

Figure~\ref{fig:3} suggests that the strength of frustration should affect the
range of the linear response regime.
The linear response regime is shorter for $J_2=0.2$ than for $J_2=1.0$,
while the non-perturbative regime is larger for $J_2=0.2$ than for
$J_2=1.0$. In fact, when $J_2=0$ (i.e. the integrable case), 
$D\propto (B_0 \omega)^{\beta}$ with $\beta=1$ for almost all the data
in the same range of $B_0\omega$ as that of Fig.~\ref{fig:3}.    

\section{\label{sec:compare} Oscillation Of energy diffusion in 
weakly-frustrated cases}

\begin{figure}
\includegraphics[width=8cm]{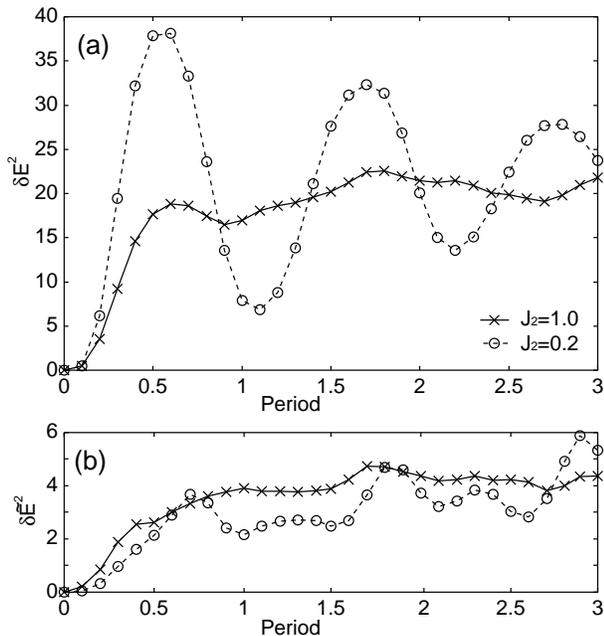}
\caption{\label{fig:4} Examples for time evolution of energy variances :  
(a) $\delta E(t)^2$ and (b)  $\delta \tilde{E}(t)^2$ (see text).
Solid lines are for $J_2=1.0$; Broken lines, $J_2=0.2$.
The parameters are
the following: $L=10$, $J_1=1.0$, $\Delta=0.3$, $B_0=1.5$, $\omega=0.5$.}
\end{figure}

We shall now proceed to investigate
oscillations of diffusion which occur in the non-perturbative regime of
a weakly-frustrated case.
Figure~\ref{fig:4}(a) shows an example of oscillatory diffusion for
$J_2=0.2$, which is compared with a non-oscillatory
diffusion for $J_2=1.0$. 
The two examples have the same set of parameters except for $J_2$.
However, the cases of $J_2=1.0$ and $J_2=0.2$ are  in the linear response
regime  and  in the non-perturbative regime, respectively.
The variance for both cases
shows normal diffusion at the very early stage of time evolution.
For $J_2=1.0$, the energy variance seems to saturate after a normal
diffusion time. On the contrary, the energy variance for $J_2=0.2$ shows
large-amplitude oscillations.
To investigate more details, we introduce another definition of energy 
variance:
\begin{equation}
   \delta \tilde{E}(t)^2 =\langle \psi (t)|[H(t)-\langle\psi (t) |
H(t)|\psi(t)\rangle]^2
   |\psi (t)\rangle .
\end{equation}
This follows a standard definition of the variance and quantifies
the degree of energy diffusion around the \textit{time-dependent
expectation} of the energy Hamiltonian.
The time evolutions of $\delta \tilde{E}(t)^2$  
corresponding to that of $\delta E(t)^2$ are shown
in Fig.~\ref{fig:4}(b). 
In the fully-frustrated case ($J_2=1.0$), the profile of 
$\delta\tilde{E}(t)^2$ is similar to that of $\delta E(t)^2$.
This observation indicates that an occupation probability spread
over the whole levels after normal diffusion of energy.

On the contrary, in a weakly-frustrated case ($J_2=0.2$) 
in Fig.~\ref{fig:4}, $\delta \tilde{E}(t)^2$
shows small-amplitude oscillations reflecting
the large-amplitude oscillations of $\delta E(t)^2$.
Most part of $\delta \tilde{E}(t)^2$ for $J_2=0.2$ is smaller than that
for  $J_2=1.0$. Furthermore, minima of $\delta \tilde{E}(t)^2$ come just
before minima and maxima of $\delta E(t)^2$.  
These observations indicates the following: an occupation probability,
which is diffusing slowly,
clustering around the expectation of energy oscillates together with the
expectation in the energy space.
To make the picture of such behavior clearer, let us
consider an occupation probability described by
\begin{equation}
  P_t(E_n)=| \langle \phi_n | \psi(t) \rangle |^2,
\end{equation}
where $|\phi_n\rangle$ is the $n$th excited eigenstate of $H_0$:
\begin{equation}
  H_0 |\phi_n\rangle =E_n |\phi_n\rangle .
\end{equation}
When $t=0$ , $P_t(E_n)$ is given by the Kronecker delta:
$P_0(E_n)=\delta_{E_n,E_0}$, where $E_0$ is the energy of the ground
state. As $t$ increases, $P_t(E_n)$ forms a wave packet in energy space
and moves to
higher levels. When the wave packet reaches some highest level, it
reflects like a soliton and moves back to lower levels. Such behavior is
repeated, although the wave packet of $P_t(E_n)$ broadens slowly.
We have actually watched this soliton-like behavior of $P_t(E_n)$ in
a form of  an animation.

\begin{figure}
\includegraphics[width=8cm]{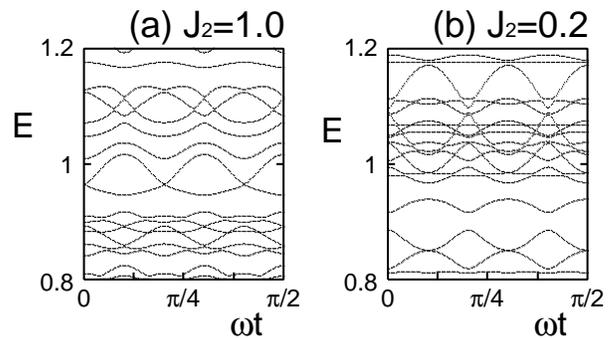}
\caption{\label{fig:5} Parts of energy spectra depending on
 adiabatically fixed time $t$ with
   $0\le t \le T/4$. Effective period is $\omega T'=2\pi/10$. The
    parameters are the following: $L=10$, $J_1=1.0$, $\Delta=0.3$, $B_0=0.8$;
   (a) $J_2=1.0$, (b) $J_2=0.2$.}
\end{figure}

The picture discussed above is also supported by the adiabatic energy
spectra in Fig.~\ref{fig:5}. 
Figures~\ref{fig:5}(a) and \ref{fig:5}(b) correspond to fully- and
weakly-frustrated cases, respectively.
Much more sharp avoided crossings appear in
Fig.~\ref{fig:5}(b) than Fig.~\ref{fig:5}(a). Some energy levels appear
to be crossing, although they are very close and never crossing in
fact. At a sharp-avoided-crossing point,
Landau-Zener formula for two adjacent levels is applicable.
Then the nonadiabatic transition leads to one-way transfer of a
population from a level to its partner, failing to result in the energy
diffusion. 
For small-$J_2$, therefore,
successive sharp avoided crossings can suppress diffusion of energy.

We believe that large-amplitude oscillations of $\delta E(t)^2$ should
be one of characteristic features of the non-perturbative regime in this
finite frustrated spin system.
In fact, similar oscillations of energy variance are seen for large
$\omega$ and large $B_0$ even when $J_2=1.0$ though the energy variance
rapidly converges after one or two periods. How long such oscillations
continue should depend mainly on $J_2$.

\begin{figure}
\includegraphics[width=8cm]{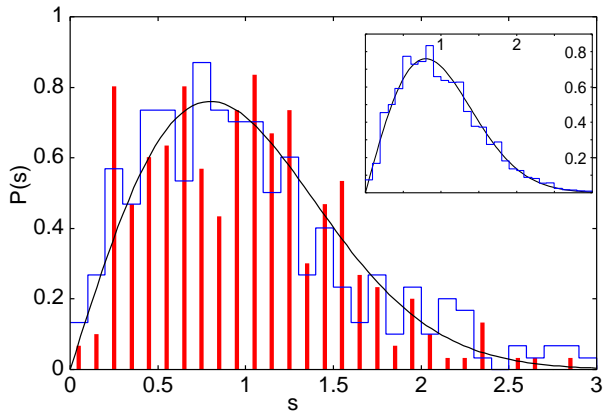}
\caption{\label{fig:6} (Color online) Level-spacing distributions at $t=\pi/4$ 
for lowest 300 levels from
  the ground state (about 10\% of all 3003 levels).  Blue histogram is for
  $J_2=1.0$; Red bars, $J_2=0.2$; Solid curve, GOE spectral statistics.
The other
    parameters are the following: $L=14$, $S^z_{\rm tot}=1$, $J_1=1.0$,
 $\Delta=0.3$,  $B_0=0.8$.
The inset is for all levels when $J_2=1.0$. The numerical methods to
 obtain the level-spacing distributions are referred in
 Refs.~\onlinecite{Kudo03,Kudo04}. }
\end{figure}

It is a notable fact that, common to both $J_2=1.0$ and $J_2=0.2$, the
level-spacing distributions in Fig.~\ref{fig:6} show GOE behavior. This
GOE behavior in the adiabatic energy spectra appears for an arbitrary
fixed time except for special points
such as $t=T=2\pi/\omega$. This fact suggests that dynamics can reveal
some various generic features of quantum many-body systems
 which can never be explained by level
statistics. The level-spacing distributions in Fig.~\ref{fig:6} convey
another crucial fact: they have been
calculated for low energy levels because our interest is in the low
energy region around the ground state. We have confirmed that the
level-spacing distributions for all energy levels in the inset is also described by
GOE spectral statistics. It is typical of this frustrated spin system that GOE level
statistics is observed already in the low energy region.~\cite{Kudo04}  

\section{\label{sec:conc}Conclusions}

We have explored the energy diffusion from the ground state
in frustrated quantum $XXZ$  spin chains under the applied oscillating
magnetic field.
In a wide parameter region of next-nearest-neighbor (NNN) coupling $J_2$
and exchange anisotropy $\Delta$,
the diffusion is normal in the early stage of diffusion.
Diffusion coefficients $D$ obey the power law
with respect to
both the field strength and driving frequency with its power being two in  the
linear response regime and equal to unity in the
non-perturbative regime.
In the case of weakened frustrations with small-$J_2$ 
we find oscillation of energy diffusion,
which is attributed to a
non-diffusive and ballistic nature of
the underlying energy diffusion.
In this way, the energy diffusion reveals generic features of the
frustrated quantum spin chains, which cannot be captured by the analysis
of level statistics.

\begin{acknowledgments}
The authors would like to thank T. Deguchi.
 The present study was partially supported by Hayashi Memorial
 Foundation for Female Natural Scientists.
\end{acknowledgments}


\begin{thebibliography}{99}
\bibitem{note1} Actually, GOE behavior has not appeared in
	     Ref.~\onlinecite{Kudo03}. In Ref.~\onlinecite{Kudo04}, it
	     is resolved why GOE behavior did not appear, and GOE
	     behavior is really shown.
\bibitem{note2} More precisely, this phrase means that
	     the time scale upon which
	     the time-dependent Hamiltonian matrix elements decorrelate
	     is much larger than the time scale corresponding to the
	     typical separation of energy levels.
\end{thebibliography}
\end{document}